\documentclass[journal=ancac3,manuscript=letter]{achemso}

\usepackage[T1]{fontenc} 
\usepackage{xcolor}
\usepackage{amssymb}

\author{Zhou~Shen}
\affiliation{Department of Physics, National University of Singapore, Singapore 117551, Singapore}
\alsoaffiliation{Max Planck Institute for the Structure and Dynamics of Matter, 22761 Hamburg, Germany}
\alsoaffiliation{Center for Free-Electron Laser Science, 22761 Hamburg, Germany}
\author{Salah~Awel}
\affiliation{Center for Free-Electron Laser Science, Deutsches Elektronen-Synchrotron DESY, 22607 Hamburg, Germany}
\author{Anton~Barty}
\affiliation{Center for Free-Electron Laser Science, Deutsches Elektronen-Synchrotron DESY, 22607 Hamburg, Germany}
\author{Richard~Bean}
\affiliation{European XFEL, 22869 Schenefeld, Germany}
\author{Johan~Bielecki}
\affiliation{European XFEL, 22869 Schenefeld, Germany}
\author{Martin~Bergemann}
\affiliation{European XFEL, 22869 Schenefeld, Germany}
\author{Benedikt~J.~Daurer}
\affiliation{Center for BioImaging Sciences, National University of Singapore, Singapore 117557, Singapore}
\alsoaffiliation{Diamond Light Source, Harwell Campus, Didcot, OX11 0DE, UK}
\author{Tomas~Ekeberg}
\affiliation{Department of Cell and Molecular Biology, Uppsala University, 75124 Uppsala, Sweden}
\author{Armando~D.~Estillore}
\affiliation{Center for Free-Electron Laser Science, Deutsches Elektronen-Synchrotron DESY, 22607 Hamburg, Germany}
\author{Hans~Fangohr}
\affiliation{European XFEL, 22869 Schenefeld, Germany}
\author{Klaus~Giewekemeyer}
\affiliation{European XFEL, 22869 Schenefeld, Germany}
\author{Mark~S.~Hunter}
\affiliation{Linac Coherent Light Source, SLAC National Accelerator Laboratory, Menlo Park, California 94025, USA}
\author{Mikhail~Karnevskiy}
\affiliation{European XFEL, 22869 Schenefeld, Germany}
\author{Richard~A.~Kirian}
\affiliation{Department of Physics, Arizona State University, Tempe, Arizona 85287, USA}
\author{Henry~Kirkwood}
\affiliation{European XFEL, 22869 Schenefeld, Germany}
\author{Yoonhee~Kim}
\affiliation{European XFEL, 22869 Schenefeld, Germany}
\author{Jayanath~Koliyadu}
\affiliation{European XFEL, 22869 Schenefeld, Germany}
\author{Holger~Lange}
\affiliation{The Hamburg Center for Ultrafast Imaging, Universität Hamburg, 22761 Hamburg, Germany}
\alsoaffiliation{Institute of Physical Chemistry, Universität Hamburg, 20146 Hamburg, Germany}
\author{Romain~Letrun}
\affiliation{European XFEL, 22869 Schenefeld, Germany}
\author{Jannik~Lübke}
\affiliation{The Hamburg Center for Ultrafast Imaging, Universität Hamburg, 22761 Hamburg, Germany}
\alsoaffiliation{Center for Free-Electron Laser Science, Deutsches Elektronen-Synchrotron DESY, 22607 Hamburg, Germany}
\alsoaffiliation{Department of Physics, Universität Hamburg, 22761 Hamburg, Germany}
\author{Abhishek~Mall}
\affiliation{Max Planck Institute for the Structure and Dynamics of Matter, 22761 Hamburg, Germany}
\alsoaffiliation{Center for Free-Electron Laser Science, 22761 Hamburg, Germany}
\author{Thomas~Michelat}
\affiliation{European XFEL, 22869 Schenefeld, Germany}
\author{Andrew~J.~Morgan}
\affiliation{University of Melbourne, Physics, Melbourne, VIC 3010, Australia}
\author{Nils~Roth}
\affiliation{Center for Free-Electron Laser Science, Deutsches Elektronen-Synchrotron DESY, 22607 Hamburg, Germany}
\alsoaffiliation{Department of Physics, Universität Hamburg, 22761 Hamburg, Germany}
\author{Amit~K.~Samanta}
\affiliation{The Hamburg Center for Ultrafast Imaging, Universität Hamburg, 22761 Hamburg, Germany}
\alsoaffiliation{Center for Free-Electron Laser Science, Deutsches Elektronen-Synchrotron DESY, 22607 Hamburg, Germany}
\author{Tokushi~Sato}
\affiliation{European XFEL, 22869 Schenefeld, Germany}
\author{Marcin~Sikorski}
\affiliation{European XFEL, 22869 Schenefeld, Germany}
\author{Florian~Schulz}
\affiliation{Institute of Physical Chemistry, Universität Hamburg, 20146 Hamburg, Germany}
\author{Patrik~Vagovic}
\affiliation{Center for Free-Electron Laser Science, Deutsches Elektronen-Synchrotron DESY, 22607 Hamburg, Germany}
\alsoaffiliation{European XFEL, 22869 Schenefeld, Germany}
\author{Tamme~Wollweber}
\affiliation{Max Planck Institute for the Structure and Dynamics of Matter, 22761 Hamburg, Germany}
\alsoaffiliation{Center for Free-Electron Laser Science, 22761 Hamburg, Germany}
\alsoaffiliation{The Hamburg Center for Ultrafast Imaging, Universität Hamburg, 22761 Hamburg, Germany}
\author{Lena~Worbs}
\affiliation{Center for Free-Electron Laser Science, Deutsches Elektronen-Synchrotron DESY, 22607 Hamburg, Germany}
\alsoaffiliation{Department of Physics, Universität Hamburg, 22761 Hamburg, Germany}
\author{Paul~Lourdu~Xavier}
\affiliation{Max Planck Institute for the Structure and Dynamics of Matter, 22761 Hamburg, Germany}
\alsoaffiliation{The Hamburg Center for Ultrafast Imaging, Universität Hamburg, 22761 Hamburg, Germany}
\alsoaffiliation{Center for Free-Electron Laser Science, Deutsches Elektronen-Synchrotron DESY, 22607 Hamburg, Germany}
\author{Filipe~R.~N.~C.~Maia}
\affiliation{Department of Cell and Molecular Biology, Uppsala University, 75124 Uppsala, Sweden}
\alsoaffiliation{NERSC, Lawrence Berkeley National Laboratory, Berkeley, California 94720, USA}
\author{Daniel~A.~Horke}
\affiliation{The Hamburg Center for Ultrafast Imaging, Universität Hamburg, 22761 Hamburg, Germany}
\alsoaffiliation{Center for Free-Electron Laser Science, Deutsches Elektronen-Synchrotron DESY, 22607 Hamburg, Germany}
\alsoaffiliation{Radboud University Institute for Molecules and Materials, 6525 AJ Nijmegen, The Netherlands}
\author{Jochen~Küpper}
\affiliation{The Hamburg Center for Ultrafast Imaging, Universität Hamburg, 22761 Hamburg, Germany}
\alsoaffiliation{Center for Free-Electron Laser Science, Deutsches Elektronen-Synchrotron DESY, 22607 Hamburg, Germany}
\alsoaffiliation{Department of Physics, Universität Hamburg, 22761 Hamburg, Germany}
\alsoaffiliation{Department of Chemistry, Universität Hamburg, 20146 Hamburg, Germany}
\author{Adrian~P.~Mancuso}
\affiliation{European XFEL, 22869 Schenefeld, Germany}
\alsoaffiliation{Department of Chemistry and Physics, La Trobe Institute for Molecular Science, La Trobe University, Melbourne, VIC 3086, Australia}
\alsoaffiliation{Diamond Light Source, Harwell Campus, Didcot, OX11 0DE, UK}
\author{Henry~N.~Chapman}
\affiliation{The Hamburg Center for Ultrafast Imaging, Universität Hamburg, 22761 Hamburg, Germany}
\alsoaffiliation{Center for Free-Electron Laser Science, Deutsches Elektronen-Synchrotron DESY, 22607 Hamburg, Germany}
\alsoaffiliation{Department of Physics, Universität Hamburg, 22761 Hamburg, Germany}
\author{Kartik~Ayyer}
\affiliation{Max Planck Institute for the Structure and Dynamics of Matter, 22761 Hamburg, Germany}
\alsoaffiliation{Center for Free-Electron Laser Science, 22761 Hamburg, Germany}
\alsoaffiliation{The Hamburg Center for Ultrafast Imaging, Universität Hamburg, 22761 Hamburg, Germany}
\author{N.~Duane~Loh}
\affiliation{Center for BioImaging Sciences, National University of Singapore, Singapore 117557, Singapore}
\alsoaffiliation{Department of Physics, National University of Singapore, Singapore 117551, Singapore}
\email{duaneloh@nus.edu.sg}


\title[An \textsf{achemso} demo]{Resolving non-equilibrium shape variations amongst millions of gold nanoparticles}

\usepackage{amsfonts}
\usepackage{siunitx}
\usepackage[version=4]{mhchem}
\usepackage{commath}
\usepackage{caption}
\usepackage{subcaption}
\usepackage{cleveref}
\usepackage{graphicx}
\usepackage{nameref}

\newcommand{\img}{\mathrm{i}}
\newcommand\given[1][]{\,#1\vert\,}
\newcommand{\bbK}{\mathbb{K}}
\newcommand{\bbM}{\mathbb{M}}
\newcommand{\bvec}{\mathbf}

\newcommand\numberthis{\addtocounter{equation}{1}\tag{\theequation}}

\DeclareMathOperator*{\argmin}{arg\,min}

\usepackage{changes}
\definechangesauthor[name={Shen Zhou}, color=orange]{SZ}

\begin{document}

\tableofcontents



\begin{abstract}
Nanoparticles, exhibiting functionally relevant structural heterogeneity, are at the forefront of cutting-edge research. Now, high-throughput single-particle imaging (SPI) with x-ray free-electron lasers (XFELs) creates unprecedented opportunities for recovering the shape distributions of millions of particles that exhibit functionally relevant structural heterogeneity.
To realize this potential, three challenges have to be overcome: (1) simultaneous parametrization of structural variability in real and reciprocal spaces; (2) efficiently inferring the latent parameters of each SPI measurement;
(3) scaling up comparisons between $10^5$ structural models and $10^6$ XFEL-SPI measurements.
Here, we describe how we overcame these three challenges to resolve the non-equilibrium shape distributions within millions of gold nanoparticles imaged at the European XFEL.
These shape distributions allowed us to quantify the degree of asymmetry in these particles, discover a relatively stable `shape envelope' amongst nanoparticles, discern finite-size effects related to shape-controlling surfactants, and extrapolate nanoparticles' shapes to their idealized thermodynamic limit.
Ultimately, these demonstrations show that XFEL SPI can help transform nanoparticle shape characterization from anecdotally interesting to statistically meaningful.
\end{abstract}

Colloidal, solid-state nanoparticles have properties defined by their size and shape, making them attractive for applications ranging from the broad field of photonics and electronics to catalysis \cite{liSemiconductingQuantumDots2018,liuColloidalQuantumDot2021,shiNobleMetalNanocrystalsControlled2021}. In the case of catalysis, for example, the nanoparticle's catalytic activity strongly depends on its size and its exposed facets, which have a strong correlation with the shape.\cite{qinSurfaceCoordinationChemistry2020,liFacilePolyolRoute2008} Hence, understanding and controlling nanoparticles' structural variations is an important aspect of synthesis\cite{modenaNanoparticleCharacterizationWhat2019}.
Commonly used post-synthesis characterization techniques (like UV-VIS\cite{haissDeterminationSizeConcentration2007}, small angle x-ray scattering (SAXS)\cite{nakamuraSizeDistributionAnalysis2003}), however, mostly measure the mean of and standard deviation of the size of nanoparticles.

To directly resolve shape variations among nanoparticles, however, requires imaging many nanoparticles individually, for example, using scanning or transmission electron microscopy (SEM or TEM)\cite{shieldsNucleationControlSize2010,woehlDirectObservationAggregative2014}. 
Tomography is sometimes used, but is time-consuming and hence limited to a few nanoparticles\cite{floreaThreeDimensionalTomographicAnalyses2013}. 
Nevertheless, electron microscopy-based characterization typically numbers in the hundreds (e.g., 300-500 particles \cite{woehlDirectObservationAggregative2014}). 
Furthermore, for larger nanoparticles, multiple scattering limits such three-dimensional (3D) shape characterization. As such, shape characterization by SEM remains largely two-dimensional (2D). Furthermore, characterization by SEM and TEM suffer from orientation bias\cite{tanAddressingPreferredSpecimen2017a} since the nanoparticles are arrested on substrates for imaging.

In contrast, high-throughput single-particle imaging with intense, ultrafast, x-ray free electron lasers (XFELs)\cite{chapmanXRayFreeElectronLasers2019} can fundamentally transform how we characterize nanoparticles. 
Single particle imaging (SPI) at the European XFEL can interrogate millions of nanoparticles in a few hours\cite{ayyer3DDiffractiveImaging2021}. 
Compared to electron microscopy, XFEL SPI is less limited by multiple scattering.
Hence, XFEL diffraction patterns of single nanoparticles closely correspond to Ewald sphere sections of the particles' Fourier volume, which in turn allows the matching of 3D structure to single-particle 2D diffraction patterns\cite{tokuhisaCoarseGrainedDiffractionTemplate2020}. 
Furthermore, nanoparticles are injected at different random orientations into the XFEL interaction region, thus avoiding the orientation bias when imaging substrate-bound nanoparticles.

Resolving the 3D shape variations among a million nanoparticles can uncover statistically meaningful insights about nanoparticles' non-equilibrium synthesis pathways.
Lurking within this opportunity, however, is a formidable statistical learning challenge: to infer the hidden parameters of the measurement of each nanoparticle, such as orientation, incident photon fluence, structural class, complex phases missing from the diffraction intensities. 
This problem is typically tackled by two types of approaches. 

The first approach induces a family of statistically likely 3D structures {\emph de novo} from large numbers of SPI patterns.
Each measurement's hidden parameters are iteratively co-refined together with these induced 3D structures.
This approach uses only prior knowledge from basic scattering physics (e.g., weak phase approximation, shot-noise limited images, etc). 
Some examples in this class extend the expand-maximize-compress algorithm (EMC)\cite{lohReconstructionAlgorithmSingleparticle2009,lohCryptotomographyReconstructing3D2010,ekebergThreeDimensionalReconstructionGiant2015, ayyerDragonflyImplementationExpand2016}, to multiple structural models \cite{choHighThroughput3DEnsemble2021, ayyer3DDiffractiveImaging2021}. 
Notably, this approach recovers an over-sampled 3D diffraction volume of each 3D structure from which its corresponding real-space electron density map is recovered using computational phase retrieval\cite{elserPhaseRetrievalIterated2003}. 
However, the number of candidate 3D structures recoverable is limited ($\lesssim 100$) by the computational memory needed to store them\cite{shenDataHeterogeneitySingle2021}.

The second approach to learning each pattern's hidden parameters uses diffraction template matching, which draws heavily on structural prior knowledge about the samples. 
Template diffraction patterns, typically created from a pool of idealized models, are used to match and classify experimentally measured SPI patterns. 
This approach does not generally require phase retrieval because each template is associated with a particular real-space model. 
Template-matching approaches were used to study variations among XFEL pulses\cite{daurerExperimentalStrategiesImaging2017,lohSensingWavefrontXray2013} and recover the histogram of sizes in $>10,000$ organelles by assuming their protein shells are spheroids\cite{hantkeHighthroughputImagingHeterogeneous2014a}.
Atsushi Tokuhisa et al. proposed a template matching method for biomolecules\cite{tokuhisaCoarseGrainedDiffractionTemplate2020} using diffraction templates generated from 3D structures in molecular dynamics simulations.
However, just like the first approach, the space of possible conformations if non-parametric is again limited by memory and compute requirements.

Here, we show how particles' shape variations (beyond the mere radius of gyration) can be simultaneously and efficiently parameterized in both real and reciprocal spaces.
This simultaneous parametrization allows us to efficiently infer the latent parameters (including complex phases) of the individual SPI patterns given a pool of 3D structures.
More importantly, this parametrization allows a principled and efficient approach to proposing and evaluating upwards of $10^5$ candidate 3D structures \emph{de novo}.

The recovered distribution of shapes (and sizes) of the millions of gold nanoparticles (two ensembles with edge lengths of approximately \SI{30}{nm} and \SI{40}{nm}, respectively\cite{ayyer3DDiffractiveImaging2021}) is telling.
We quantified the degree of asymmetry in each nanoparticle from the distribution of their $(111)$ and $(100)$ facet areas.
We also discovered a relatively stable `shape envelope' in two different ensembles of nanoparticles.
Since both ensembles were extracted at different times in a common crystal growth trajectory, we could extrapolate their particle shapes to large crystals in the thermodynamic limit.
Furthermore, we found hints of finite-size effects related to the surfactant used to control the nanoparticles' shape. 
These studies demonstrate the potential of XFEL for studying nonequilibrium systems that are difficult to image directly by conventional means or too heavy for molecular dynamics simulation.

\section{Results and Discussion}

\subsection{Synthesis and measurements of nanoparticle ensembles.}\label{ssec:synthesis}
Two ensembles of truncated octahedral gold nanoparticles were synthesized using the protocols described elsewhere \cite{liFacilePolyolRoute2008,luSizetunableUniformGold2017}. 
We used a solution comprising \ce{HAuCl4} as the precursor and poly (diallyldimethylammonium) chloride (PDDA) as a surfactant. 
This mixture was introduced into a round-bottom flask containing 1,5-pentanediol (PD) solution and refluxed in an oil bath at a temperature maintained at up to \SI{225}{\celsius}.
Two ensembles of nanoparticles were created by quenching this mixture in a room-temperature water bath after approximately \SI{4}{min} (for sample oct30) or \SI{7.5}{min} (for sample oct40) of reaction time.
Following quenching, the resulting crude nanoparticle mixture underwent thorough purification to eliminate excess ligands. This purification involved sequential centrifugation steps in acetone and water. The resulting nanoparticle pellet was re-suspended in ultra-pure water, which is used for subsequent XFEL imaging. Using scanning electron microscopy (SEM) (\cref{fig:sem}) on small batches of the nanoparticles from these two ensembles, we determined their nominal average widths as \SI{30}{nm} (oct30) and \SI{40}{nm} (oct40). 

The samples were injected by an electrospray injector and focussed with an aerodynamic lens stack into the stream of XFEL pulses at the European XFEL (EuXFEL), as described in Ayyer {\it et al.} \cite{ayyer3DDiffractiveImaging2021}. 
Due to the high pulse repetition rates of the EuXFEL, approximately 105 and 65 diffractions of single particles were accumulated per second to comprise the oct30 and oct40 datasets respectively. 
Millions of such XFEL measurements were used to reconstruct two average 3D structures, each representing either the oct30 or oct40 ensembles \cite{ayyer3DDiffractiveImaging2021}. 
The widths of these two average 3D structures (for oct30 and oct40 respectively) were \SI{35}{nm} and \SI{40}{nm}; the longest edge lengths of their $(111)$ facets were \SI{20}{nm} and \SI{27}{nm}.

Two-dimensional (2D) {\it in-silico} classification\cite{ayyer3DDiffractiveImaging2021} filtered out empty shots, multiple-particle shots, and diffraction patterns likely belonging to non-octahedral nanoparticles in both oct30 and oct40 ensembles.
After post filtration, \num{1287570} and \num{823202} patterns remained in the oct30 and oct40 datasets respectively.

\subsection{Parametrizing structural variations}

Earlier analyses of the large oct30 and oct40 datasets showed \cite{ayyer3DDiffractiveImaging2021} noticeable structural variations amongst truncated octahedral nanoparticles, which led to the averaged 3-dimensional models showing `rounded' $(100)$ facets.
Our goal here is to characterize these variations in a statistically robust and meaningful way.

The space of nanoparticle structural variations, which is resolvable by our experiment, lives in a $10^5$-dimensional space (see Methods Section \nameref{dof}).
However, we seek only the \emph{posterior distribution of their first-order distortions from the average truncated octahedron}.
Such distortions can be efficiently parameterized with a simpler 42-dimensional {\it free-facet truncated octahedron (FFTO) model}, which consists of the vertex positions of 14 facets of a truncated octahedron (\cref{fig:pca}(b)).

We used a weighted Monte Carlo importance sampling scheme to sample the oct30 and oct40 ensembles' posterior distribution in the 42-dimensional FFTO space.
We then parameterized the dominant structural variations within these posterior distributions, which allowed us to infer the nanoparticles' synthesis conditions directly.

\subsection{Posterior estimation using Monte Carlo importance sampling}\label{ssec:MonteCarlo}

Exhaustively resolving the posterior probability of a nanoparticle's structure in a 42-dimensional FFTO is computationally prohibitive.
In a naive approach, this would involve comparing each diffraction pattern against a large number of possible 3-dimensional models that densely cover this 42-dimensional space.
Each comparison, in turn, requires checking the most likely orientation in which each pattern could arise within each model.
Instead, we know these structures stay close to a truncated octahedron \cite{ayyer3DDiffractiveImaging2021}, making the vast majority of these models in the FFTO space unlikely or, equivalently, {\it unimportant} in this analysis.   
Hence, a much smaller and non-uniformly spaced pool of FFTO models can capture the most important nanoparticle structures.

\begin{figure*}[htpb]
    \centering
    \includegraphics{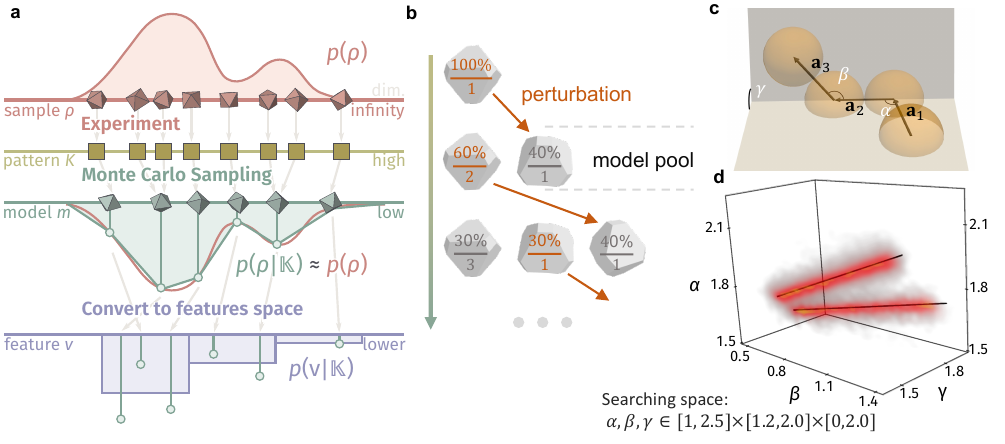}
    \caption{(a) Framework to estimate the structural posterior distribution of particles from their experimental measurements (i.e., diffraction patterns). 
    (b) The weighted Monte Carlo importance sampling scheme, which includes model selection (red numbers), perturbation (red arrows), and weights updating. (c) Rotational degrees of freedom ($\alpha, \beta, \gamma$) in our artificial ensemble of nanoparticles, each as a 4-ball chain. (d) Posterior distribution of ensemble in (c) using our Monte Carlo scheme in (a). The ground truth structural trajectory is shown with the twin black lines from which diffraction patterns are randomly generated. The pool of Monte Carlo models is rendered in a semi-transparent point cloud, where higher red intensity indicates models with higher data likelihood given the diffraction patterns.
    }
    \label{fig:framework}
\end{figure*}

To estimate the posterior distribution of likely structures, we seek a pool of FFTO models, $\bbM = \{\rho_1, \rho_2, \dots\}$, that efficiently sample the posterior space (see \cref{fig:framework}(a)). 
To paraphrase, this pool should encompass the set of FFTO models that are most likely to produce the experimentally measured oct30 and oct40 diffraction patterns, $\bbK = \{ K_1, K_2, \dots\}$.

We use a weighted Monte Carlo (MC) importance sampling scheme to efficiently accumulate this pool of models (\cref{fig:framework}(b)).
This MC model pool starts with an initial model that is randomly perturbed from the average 3D structure, which is approximated from the single-model reconstruction result of the whole dataset. 
New models in the FFTO space are iteratively added to this pool in three steps: 
(1) select a weighted random model from the existing model pool;
(2) perturb this selected model; 
(3) add the perturbed model to the pool and update all the models' weights.
For this MC scheme to sufficiently sample the FFTO space, it needs to explore the space of less likely models. 
We do this by penalizing excessive selections of the most likely models in the pool.
Hence, the weights used to select random models in the first step depend on the ratio between the following two quantities: the percentage of diffraction patterns that are likely due to each model in the pool as defined by \cref{eqn:likelihood} and shown as numerators in \cref{fig:framework}(b); 
and the number of times each model was selected for perturbation in step 2, which starts at 1 for each added model, and shown as denominators in \cref{fig:framework}(b).
The numerator ensures that we explore the neighbourhood around likely models, while the denominator favors selecting less frequently visited models.

With this pool of models, we can evaluate the posterior probability of various nanoparticle features $\nu$ (e.g., length, shape, volume, asymmetry, etc) given the diffraction measurements of the oct30 and oct40 ensembles ($\bbK$).
This probability is similar to a weighted voting scheme: 
each model in the pool ($\rho_i$) casts a vote for a particular feature, and this vote is weighted by that model's posterior probability given all measurements. 
This leads to the posterior estimates in \cref{eq:bayes,eq:pattern_dist}, which we derive in the \nameref{sec:Method} section.


We demonstrate this framework on an artificial ensemble of flexible particles.
Each particle consists of four identical balls that are sequentially attached (\cref{fig:framework}(c)). 
Each particle's structure is described by three angles defined in their body axes (bond angles $\alpha$, and $\beta$; dihedral angle $\gamma$).
All possible particle structures are confined to a ground truth linear trajectory (black line in \cref{fig:framework}(c)).
Since the four balls in each particle are identical, swapping the first and last balls, which also swaps $\alpha$ and $\beta$, yields identical diffraction patterns.
This leads to a duplicate of the ground truth trajectory in ($\alpha$, $\beta$, and $\gamma$) space.
From  \num{100000} diffraction patterns of randomly rotated particles with random structures along this ground truth trajectory, we correctly reconstructed the posterior distribution of structures shown in red in \cref{fig:framework}(d).
Details about this artificial ensemble are discussed in the \nameref{sec:Method} section.

Validating our estimated posterior distribution is important, especially when the raw data is sparse and incomplete.
Since we did not have the ground truth posterior for the oct30 and oct40 datasets for validation, we checked that our estimated posterior has converged and is self-consistent.
Briefly, we used the reconstructed model pools as a `proxy' to the ground truth to generate random test diffraction patterns. 
These generated test patterns were then used to accumulate a second pool of models, which were compared to the ground truth `proxy' for repeatability. 
Details are described in \nameref{sec:Method}.

\subsection{Dominant structural modes in nanoparticle ensembles}\label{subsec:identify}
\begin{figure*}[htpb]
    \centering
    \includegraphics{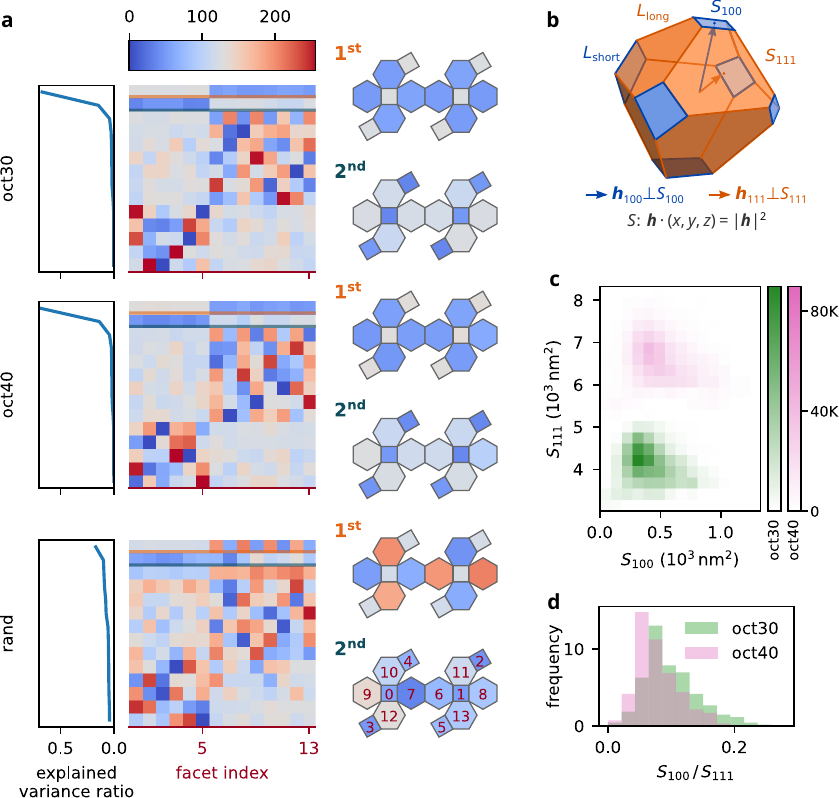}
    \caption{Quantifying the dominant modes of variation in nanoparticle ensembles 
    (a) Principal Component Analysis (PCA) was applied to the \emph{facet-area} features of three nanoparticle ensembles: oct30, oct40, and a randomly generated synthetic one.
    The PCA modes of these features are shown as rows of the matrices (middle block, linear color scale from blues to reds as negative to positive);  these modes (i.e., rows) are sorted by their explained variance ratio (left block). In the right block, we see the corresponding net plots of the two most dominant PCA modes of each ensemble, where the facets are colored according to the modal variations. The facet indices of the PCA mode columns are laid out on the bottom net plot.   
        (b) The coordinate system for our FFTO models, where the total areas of the $(111)$ and $(100)$ facets, denoted as $S_{111}$ and $S_{100}$, are shown in orange and blue respectively.
        (c) The facet-area features of oct30 and oct40 are projected onto the $S_{111}$-$S_{100}$ subspace.  
        (d) The distribution of the $S_{111}/S_{100}$ area ratio for oct30 and oct40 demonstrates that the former experiences a more significant truncation along the $(100)$ directions.
}%
    \label{fig:pca}
\end{figure*}

Our weighted Monte Carlo importance sampling of the nanoparticles' posterior distribution yielded a pool of FFTO models representing the most probable nanoparticle structures in our oct30 and oct40 diffraction datasets.
However, each FFTO model is described by a 42-element vector, which \emph{still} has far too many dimensions for us to visualize. 

Fortunately, these 42 numbers are not mutually independent, as they can describe the same nanoparticle structure but at a different orientation and/or translation.
Hence, dimensionality reduction should be possible.
To accomplish this, we mapped each FFTO model into a \emph{facet-area representation}, which consists of an ordered list of the areas of each model's 14 facets.
In this representation, the areas of the $(100)$ direction facets are indexed from $0$ to $5$, and those of the $(111)$ direction facets are indexed from $6$ to $13$ (\cref{fig:pca}(a)).
This set of facet-areas is not only invariant under rotations and translations but is also conveniently related to each model's surface free energy \cite{modenaNanoparticleCharacterizationWhat2019}.

To simplify our analysis of the estimated posterior, we `hard-assigned' each diffraction pattern $K$ only to its most probable model in the Monte Carlo FFTO model pool.
Each time an FFTO model is deemed most likely for a pattern, we projected this FFTO model into its 14-dimensional facet-area feature space and then appended this facet-area model to a growing list.
Since an FFTO model might be deemed most likely by multiple diffraction patterns, this model's facet-area features might appear multiple times within this list.
For brevity, we will refer to this list of features as the \emph{facet-area point cloud}, or equivalently, the $\envert{\bbK}\times 14$ matrix $X$.

Due to the octahedral symmetry of our models, the order of these 14 numbers in a facet-area feature can be changed by applying any rotation operation within this symmetry group. This redundancy is eliminated (details in \nameref{sec:Method} section) since we are not interested in orientational differences amongst nanoparticles.

The primary structural variations manifest in this facet-area point cloud ($X_\text{oct30}$ or $X_\text{oct40}$), which has been reduced in symmetry, are examined using principal components analysis (PCA) (\cref{fig:pca}(a)). 
To do this, we decomposed $X$ into 14 modes, sorted by their explained variances. 
Modes with higher explained variance describe more frequent structural variations.
We color the modes of these facet-area variations in \cref{fig:pca}(a) for both $X_\text{oct30}$ and $X_\text{oct40}$.
For comparison, we include an ensemble of randomly perturbed truncated octahedra $X_\text{rand}$ with \num{10000} points.
Each point was perturbed from an average canonical FFTO model whose facets are aligned perfectly along the $(111)$ or $(100)$ directions.
The first two PCA modes of $X_\text{oct30}$, $X_\text{oct40}$, and $X_\text{rand}$ are colored in the same manner in their accompanying polyhedral net plot.

These dominant facet-area PCA variations reveal that the surface energy densities of the nanoparticles' $(111)$ and $(100)$ facets are distinct.
More than 80\% of the facet-area variations of the \emph{millions of nanoparticles} in $X_\text{oct30}$ and $X_\text{oct40}$ can be explained by their respective first two PCA modes. 
The most dominant PCA mode shows that the $(111)$ facet-areas tend to be correlated, while the next mode shows similar correlations amongst the $(100)$ facet-areas.
This correlation is notably absent in $X_\text{rand}$, where no constraints were imposed on the ratios amongst the surface energy densities of different facets.
The correlations within the first two modes can be explained by the fact that the free energy of a nanoparticle includes the terms $\gamma_{111}S_{111}$  and  $\gamma_{100}S_{100}$, where $\gamma$ and $S$ denote the surface energy densities and total areas of the subscripted facets.
Our observed correlations are hence consistent with expectation that $\gamma_{111}$ and $\gamma_{100}$ are different for these octahedral nanoparticles\cite{liFacilePolyolRoute2008,luSizetunableUniformGold2017}.

Relatedly, the variations in the $(111)$ facet areas are approximately four times higher than those of $(100)$ facets.
This indicates that much of the changes in the surface area of oct30 and oct40 nanoparticles still lie on their $(111)$ facets.

The third-ranked dominant PCA modes of $X_\text{oct30}$ and $X_\text{oct40}$ in \cref{fig:pca}(a) are similar, but likely due to random fluctuations since they resemble the top-ranked mode for $X_{\text{rand}}$.
The alternating signature of this mode is largely due to eliminating symmetries in these features, which was also performed on $X_{\text{rand}}$.

 These observations quantify the degree to which each nanoparticle's structural variations are highly correlated to the areas of their $(111)$ and $(100)$ facets.
 Furthermore, since the areas of the $(111)$ and $(100)$ facets are separately correlated, these variations can be further reduced to just the two-dimensional space of $S_{111}$ vs $S_{100}$ (the sum of the $(111)$ and $(100)$ facet-areas respectively). 
We project the posterior distributions for $X_\text{oct30}$ and $X_\text{oct40}$ into the $S_{111}$-$S_{100}$ subspace in
\cref{fig:pca}(c). 

Finally, the $S_{100}/S_{111}$ ratio of each FFTO model is proportional to the extent  of truncation along the $(100)$ octahedral facet, where smaller ratios indicate less truncation.
By projecting the posterior distribution into the $S_{100}/S_{111}$ subspace in \cref{fig:pca}(d), we see that the oct40 is less truncated than the oct30 ensemble. 
In the size range of our experiment (\SI{30}{nm} to \SI{50}{nm}), smaller particles exhibit a tendency towards being more spherical. A similar behavior was observed in decahedral multiply twinned gold NPs\cite{alpayAreNanoparticleCorners2015}. 


\subsection{Evidence of non-equilibrium growth from posterior distributions}

\begin{figure}[htpb]
    \centering
    \includegraphics[]{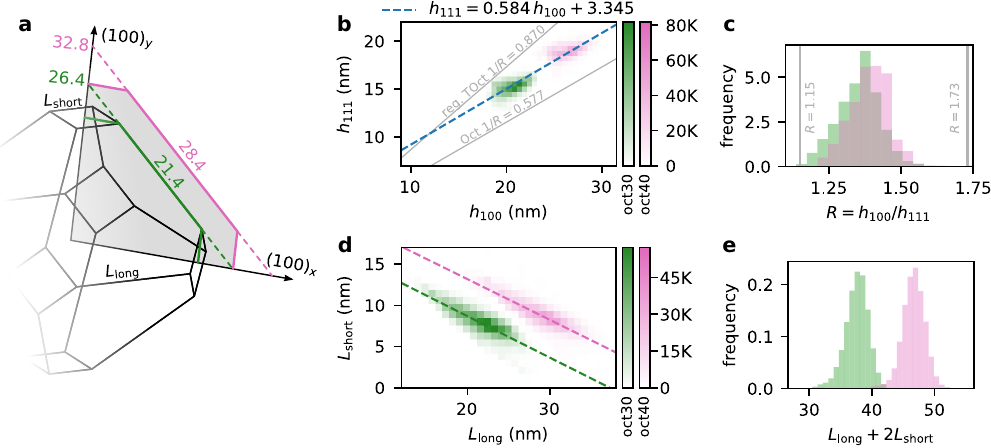}
    \caption{Signs of non-equilibrium growth in truncated octahedra.
    (a) A truncated octahedron (black edges) shown with the oct30 and oct40 octahedral envelopes (green and pink dashed lines).
    The distances of the $h_{100}$ vertices in the oct30 and oct40 envelopes are \SI{26.4}{nm} and \SI{32.8}{nm} respectively.
    The average longest edge lengths of the oct30 and oct40 $(111)$ facets are \SI{21.4}{nm} and \SI{28.4}{nm} respectively.          
    (b) The posterior distributions of $h_{111}$ vs $h_{100}$ for oct30 and oct40 denote the average distances between the origin and the $(111)$ and $(100)$ facets respectively.
    Gray lines show the $(h_{111}, h_{100})$ relationship for an equilibrium regular octahedron and a regular truncated octahedron.
    Blue dotted line interpolates between the oct30 and oct40 posterior distributions (fit function as plot title).
    (c) Frequency histogram of $R=h_{111}/h_{100}$ projected from (b), annotated with the ratios of a regular truncated octahedron ($0.577$) and an octahedron ($0.870$). 
    (d) The distributions of the average lengths of the shorter and longer edges of the truncated octahedra in oct30 and oct40 ($L_\text{long}$ vs $L_\text{short}$), each fitted to a line.
    Both lines fit for $-0.5$ slope, indicating most nanoparticles are constrained to an octahedral envelope of edge length (i.e., $L_\text{long}+ 2 L_\text{short}$) of \SI{26.4}{nm} (oct30), or \SI{32.8}{nm} (oct40). 
    (e) The frequency-distribution of the envelopes' edge length, $L_\text{long}+ 2 L_\text{short}$, show \SI{\leq 3}{nm} FWHM deviation in both oct30 and oct40.}%
    \label{fig:linear}
\end{figure}

The PCA of the posterior distributions in \cref{fig:pca} show that the first-order structural variations in either oct30 or oct40 can be further reduced to features associated with either each nanoparticle's $(111)$ facets or those with their $(100)$ facets.
Here are two possible feature pairs that can be physically interpreted.
The first pair we chose is $(h_{100}, h_{111})$: the average distances of its $(100)$ and $(111)$ facets from each nanoparticle's origin respectively. 
These distances are key parameters in the Wulff construction used to describe the equilibrium shapes of crystals.
The second pair of features is $(L_\text{short}, L_\text{long})$, which are the average lengths of two types of edges: twenty-four shorter edges of $(100)$ facets (blue edges in \cref{fig:pca}(b)), and the remaining twelve longer edges (orange edges in \cref{fig:pca}(b)) respectively. 

We can gain valuable insights into the overall growth trajectory of both nanoparticle ensembles  by extrapolating from and interpolating between the $(h_{100}, h_{111})$ features of the oct30 and oct40 ensembles (\cref{fig:linear}(b)).
According to the Gibbs-Wulff theorem, when a constant volume crystal attains its equilibrium shape, the ratio $R=h_{100}/h_{111}$ equals the
ratio between the surface tensions of its $(100)$ and $(111)$ facets, denoted as $\gamma_{100}/\gamma_{111}$. In our specific case, density functional theory\cite{vitosSurfaceEnergyMetals1998} predicts that $R_0=1.27$. Additionally, the ideal (untruncated) octahedron and regular truncated octahedron exhibit $R$ values of $\sqrt{3}$ and $\sqrt{4/3}$, respectively.

In \cref{fig:linear}(b), the posterior distributions of oct30 and oct40, when projected to the $(h_{100}, h_{111})$ subspace, fit the dashed blue line given by $h_{111} = 0.584 h_{100} + 3.345$.
Since the only difference between the synthesis of the oct30 and oct40 ensembles is their reaction times\cite{luSizetunableUniformGold2017,ayyer3DDiffractiveImaging2021}, we assume that these two ensembles are two ``snapshots'' of the same crystal growth trajectory connected by this fitted blue line. 

If we extrapolate this fitted growth trajectory forward in time, assuming the nanocrystals could grow towards their thermodynamic limit (i.e., $h_{100}\to \infty$)\cite[Chapter~13]{vanselowChemistryPhysicsSolid1988} it would approach the facet displacement ratio of $R=h_{100}/h_{111}\to 1.71 $.
This ratio is smaller than $R=\sqrt{3} \approx 1.73$ of a regular (untruncated) octahedron.
This trend suggests that larger crystals beyond those in oct40 will always exhibit some truncation on their $(100)$ facets. This is consistent with the larger octahedra synthesized by Lu {\it et al.} that show `rounded' $(100)$ facets \cite{luSizetunableUniformGold2017}.
Extrapolating this growth trajectory backward in time, it intersects the regular truncated octahedra ratio of $R=\sqrt{\frac{4}{3}}\approx 1.15$ when $h_{100}$ is around \SI{11}{nm}, where the particles are most spherically symmetric.
  

Both nanoparticle ensembles in \cref{fig:linear}(c) deviate significantly from the reference $R_0$. The oct40 ensemble ($=1.42$) deviates more prominently than the oct30 ensemble ($=1.34$).
This deviation can be attributed to the synthesis process of these nanocrystals, wherein the cationic surfactant PDDA was employed to inhibit the growth of $(111)$ facets.
The presence of adsorbed PDDA molecules impedes the contact between the crystal facets and free gold atoms in the solution. It preferentially attaches to the $(111)$ facets rather than the $(100)$ facets, this results in an increased value of the sample's $R=h_{100}/h_{111}$ compared to the $R_0$.

In the \nameref{sec:Method} section, we propose a phenomenological model that shows how the adsorption efficiency of PDDA might change as the nanocrystals grow due to finite-area effects.
Consider how each PDDA polymer (about \SI{400}{kDa} to \SI{500}{kDa}) is a linear molecule that spans several nanometers, which is comparable to the sizes of small nanocrystals.
We assume that at our elevated nanocrystal growth temperatures, each PDDA polymer must be securely adsorbed onto a crystal facet via a minimum number of van der Waals contacts, $N_{\text{min}}$.
Hence, the attachment of an incoming PDDA polymer to a crystal facet will be frustrated by adsorbed PDDA that occupy possible attachment sites (\cref{fig:coverage}(a)).
When the area of a facet that is covered by randomly adsorbed PDDA polymers reaches a critical fraction, the average number of contiguous attachment sites available to an incoming PDDA molecule falls below $N_{\text{min}}$. 
Consequently, the attachment rate of new PDDA molecules slows dramatically due to frustrated attachment.
Our simple model shows that this critical area fraction is reached sooner for smaller facet areas due to the size of randomly adsorbed PDDA.
Conversely, this finite-area effect will become unimportant when the crystals are much larger than the average size of the PDDA molecule. 

The finite-area effect mentioned in the previous paragraph suppresses both $\gamma_{100}$ and $\gamma_{111}$.
Recall that the Gibbs-Wulff theorem states that in an ideal crystal of constant volume at equilibrium, $\gamma \propto h$.
Since $R=h_{100}/h_{111} = \gamma_{100}/\gamma_{111}$, one might expect $R$ to be constant for such idealized nanocrystals.
However, as shown in \cref{fig:pca}(c), we observe that the area of the $(111)$ facets expands relative to the $(100)$ facets as the crystals grow from oct30 to oct40.
Therefore, the finite-area effect suppresses $\gamma_{111}$ less than $\gamma_{100}$ in oct40 compared to oct30.
This leads to $R$ increasing in \cref{fig:linear}(c) from oct30 to oct40.

The posterior distributions of oct30 and oct40 in the $(L_\text{short}, L_\text{long})$ subspace of \cref{fig:linear}(d) show a linear trend with a $-\frac{1}{2}$ slope, which means $L_\text{long} + 2 L_\text{short}$ is close to a constant.
From \cref{fig:linear}(a) we see that $L_\text{long} + 2 L_\text{short}$ is the edge length of the enveloping octahedron. 
Hence, the distribution in \cref{fig:linear}(d) for the millions of nanoparticles reveals two notable insights: a separate octahedral envelope that encompasses the structural variations within each ensemble; and the relative truncation of the $(100)$ facets within this envelope decreases from oct30 to oct40.

The edge lengths of the enveloping octahedra for oct30 and oct40 are calculated from linear fits to their projected posterior distributions in \cref{fig:linear}(d).
Consistent with the explained variance ratios of the PCA in \cref{fig:pca}(a), the nanoparticles' shape variations within oct30 or oct40 are largely due to different extents of $(100)$ facet truncations within each ensemble's octahedral envelope.
The $h_{(100)}$ distances of oct30 and oct40 octahedral envelopes increased from \SI{26.4}{nm} to \SI{32.8}{nm} (\cref{fig:linear}(a)) respectively, while their corresponding edge lengths increased from \SI{37.3}{nm} to \SI{46.4}{nm} (\cref{fig:linear}(e)). Notably, the average longest edge lengths of the $(111)$ facets for oct30 and oct40 in \cref{fig:linear}(a) are \SI{21.3\pm 2.8}{nm} and \SI{28.4\pm 3.3}{nm} respectively.
These lengths overlap with those from the average 3-dimensional models reconstructed in Ayyer {\it et al.} \cite{ayyer3DDiffractiveImaging2021} (see Section on \nameref{ssec:synthesis}).


\section{Conclusions}
 
Megahertz XFEL sources offer tremendous potential for inferring properties of particle ensembles numbering in the millions.
The posterior distributions of these large ensembles of nanoparticles detail their structural dynamics and interactions. 
However, estimating these distributions is a computationally expensive and data-intensive endeavour.

This paper describes a scalable Monte Carlo importance-sampling framework to robustly estimate the posterior distribution of structural variations amongst very large numbers of single nanoparticles.
By explicitly parameterizing these structures in the free-facet truncated octahedra (FFTO) space, we were able to avoid ambiguous features that often arise in prior-free induced `manifolds' on XFEL datasets.
Additionally, we also propose methods to validate the consistency of the recovered posterior distributions that circumvent the issues with Fourier Shell Correlation which is typically used in XFEL single particle imaging\cite{shenEncryptionDecryptionFramework2021}.  

Our manuscript also details practical implementation strategies to accelerate this importance sampling for millions of noisy and incomplete single-particle XFEL diffraction patterns.
This includes an analytical approach to directly compute the diffraction pattern from polyhedra that can be efficiently implemented on GPGPUs (\nameref{sec:supplementry}). 
These strategies allow us to infer structural heterogeneity from datasets that are at least two orders of magnitude larger than what was previously attempted for single-particle XFEL imaging.

We interpreted such uncommonly high-dimensional posterior distributions using PCA, which showed that the structural variations within our truncated octahedra ensembles can be described by two independent degrees of freedom.
By picking different projections of these two degrees of freedom, we inferred key signatures of non-equilibrium growth dynamics of nanocrystal growth, which led us to hypothesize a finite-area effect that might drive these dynamics away from equilibrium.

Our work shows a scalable statistical learning path to posterior estimation on massive datasets in high-throughput XFEL facilities worldwide. 
More broadly, it illuminates a similar path for data-driven heterogeneity mapping in single particle imaging, including cryo-electron microscopy.
The four-ball model example in our manuscript shows that our framework also works for flexible particle chains (e.g., polymers, polypetides, etc).
Here, efficiently parameterizing an object's structure is critical. 
Since the free-energy landscape of biomolecules can be embedded in a low-dimensional surface \cite{neupaneProteinFoldingTrajectories2016}, a low-dimensional parameterization of their structures might be possible.
Ultimately, we have both the datasets and statistical learning tools for an unprecedented window into the hidden and chaotic world of nanoparticle dynamics.

\section{Acknowledgements}
The authors acknowledge the support of John~C.H.~Spence, who continues to inspire many of us. N.D.L.\ would like to thank the support of the Early Career Grant by the National University of Singapore, and the Singapore Ministry of Education AcRF Tier 1 grant. Z.S.\ thanks the Ph.D.\ scholarship from the Physics Department at the National University of Singapore. Both N.D.L.\ and Z.S.\ are grateful for the support by Bai~Chang from the Centre for Bio-imaging Sciences at the National University of Singapore for IT infrastructure support. Paul~Lourdu~Xavier acknowledges a fellowship from the Joachim Herz Stiftung, Germany. Paul~Lourdu~Xavier and Henry~N.~Chapman acknowledge support from the Human Frontiers Science Program, France (grant No. RGP0010/2017). Richard~A.~Kirian would like to thank the support of the National Science Foundation, BioXFEL Science and Technology Center (award \#1231306) and the National Science Foundation, Directorate for Biological Sciences (award \#1943448).

\section{Methods} \label{sec:Method}

\subsection{Degrees of freedom in nanoparticles' structure} \label{dof}
Upon inspecting the 2D class averages of oct40 nanoparticles (\cref{fig:2demc}), it is observed that most diffraction patterns are characterized by approximately 12-15 radial resolution elements, as defined by Loh and Elser\cite{lohReconstructionAlgorithmSingleparticle2009}.
Consequently, the electron density maps of each nanoparticle can be represented by a 3D grid containing approximately $10^5$ resolution elements, calculated from $\sim (2\times 15 + 1)^3$.
Although it is possible to determine the modal structures \cite{choHighThroughput3DEnsemble2021} of our nanoparticle ensemble in this $10^5$-dimensional space, efficiency can be significantly enhanced with prior knowledge about these variations.

\subsection{Data likelihood model}\label{subsec:basic_EMC}

Our aim is to infer the posterior distribution $p(\rho\given \bbK)$, representing the structural conformations ($\rho$s) within an ensemble, using collected diffraction patterns ($\bbK$). However, directly applying Bayes' theorem, $p(\rho \given \bbK) p(\bbK) = p(\bbK \given \rho) p(\rho)$, to estimate $p(\rho\given\bbK)$ is not feasible due to the imprecision in defining both terms on the right-hand side. Conformation $\rho$ is conceptualized as a function that assigns electron densities to points in real space, indicating that $\rho$'s domain is infinitely dimensional. Defining $p(\rho)$ on such domain is a significant challenge. Furthermore, since observed patterns are derived from different instances of $\rho$, for any specific pair of $K$ and $\rho$, $p(K\given\rho)$ is likely to be zero. This leads to $p(\bbK \given \rho)$ frequently approaching zero, causing the formula to be ill-defined and computationally unstable.

Instead of studying the full dataset, we should focus on a single pattern. 
In the context of a specific pattern pattern $K$ and a particular feature $\nu$, the 
posterior probability, $p(\nu\given K)$, essentially quantifies how much $K$ is distributed or ``voted'' to a $\nu$.
Then the averaged $p(\nu\given K)$ over $K\in\bbK$, 
$p(\nu\given\bbK) \equiv \bigl\langle p(\nu\given K)\bigr\rangle_{K\in\bbK}$,
gives us an overall posterior estimation over a whole ensemble.

According to Bayes' theorem:
\begin{equation}
    p(\nu \given K) p(K) = p(K \given \nu) p(\nu) \; .
    \label{eq:bayes}
\end{equation}
To make progress here, we will need an uninformative-prior assumption about the feature space: $p(\nu)$ is a constant. In addition, the value of $p(K\given \nu)$ is approximated by $p(K\given \nu; \bbM)\equiv\sum_{\rho\in \bbM}p(K\given \rho)p(\rho \given \nu)$. Assuming the uninformative prior, $p(\rho \given \nu)=1 / N_{\nu, \bbM}$, where $N_{\nu, \bbM}$ is the number models in $\bbM$ having feature $\nu$.
The definition of $p(K\given\rho)$ will be discussed later in \cref{eqn:likelihood}. To summarize the discussion above with formulae, we have
\begin{equation}
    \begin{aligned}
        p(\nu\given K)&\approx p(\nu\given K; \bbM)\\
        &\propto p(K\given \nu; \bbM)\\
        &=\sum_{\rho\in\bbM} p(K\given \rho)p(\rho\given\nu)\;,
    \end{aligned}
\end{equation}
and
\begin{equation}
    p(\nu\given\bbK; \bbM) \equiv \frac{1}{|\bbK|}D(\nu\given\bbK; \bbM)\;,
    \label{eq:pnubbKbbM}
\end{equation}
where
\begin{equation}
    D(\nu\given\bbK; \bbM)\equiv \sum_{K\in\bbK}p(\nu\given K)\;.
    \label{eq:pattern_dist}
\end{equation}
It is worth to notice that $\sum_{\nu\in V} D(\nu\given \bbK; \bbM)=|\bbK|$.



In XFEL-SPI, numerous far-field diffraction patterns ($\bbK$) are captured, each originating from a distinct particle in the ensemble illuminated by a single x-ray pulse. Occasionally, multiple particles may diffract from a single pulse, but this is predominantly filtered out {\it in silico} (as explained below).
Disregarding background and inelastic scattering, these patterns represent the far-field diffraction resulting from the phase shift induced on the x-ray pulses by a particle's two-dimensional (2D) projected scattering potential.
The orientation of each particle is unmeasured and has to be inferred \cite{lohReconstructionAlgorithmSingleparticle2009}.
Due to the photon limitation, these patterns essentially represent the Poisson-sampled Ewald sphere tomograms of the target particle's three-dimensional (3D) diffraction intensity $W$.
This diffraction intensity varies linearly with the unmeasured local fluence of the XFEL pulse that illuminated each particle.
Taken together, the likelihood\cite{ayyerDragonflyImplementationExpand2016} of measuring a particular pattern $K$ given a tomogram $W_{Q}$ of the particle presented at orientation $Q$ is
\begin{equation}
     p(K\given Q,W, \phi ) = \prod_{t \in \text{detector}} \frac{ \text{e}^{-\phi_K W_{Q t}} \,\bigl(\phi_K
     W_{Q t}\bigr)^{K_{t}} }{K_{t}!}\text{,}  \label{eqn:likelihood}
\end{equation}
where $t$ indexes the detector's pixels, and $\phi_K$ is the local fluence rescaling factor for $K$\cite{lohCryptotomographyReconstructing3D2010}.

For a weakly scattering particle $\rho$, its diffraction intensities $W$ are the squared modulus of the Fourier transform of the particle's real-space electron density distribution $\rho(\mathbf{r})$, which is represented as $W(\mathbf{q})=\envert[2]{\mathcal{F}_{\mathbf{r}\to\mathbf{q}}[\rho(\mathbf{r})]}^2$.
Thus, the likelihood of measuring a pattern $K$ given an electron density $\rho$ is
\begin{equation}
    p(K\given \rho) \equiv p(K \given W) = \int d \phi_K \sum_{Q \in \mathbb{Q}} p(K\given Q,W, \phi_K)\; p(Q) p(\phi_K) \; ,
    \label{eq:pkrho}
\end{equation}
where $\mathbb{Q}$ is the set of orientations in $SO(3)$ space considered for particle $\rho$. 
The likelihood $p(K\given \rho)$ here estimates how well each pattern $K$ is matched to our Monte Carlo model $\rho$.

To simplify \cref{eq:pkrho}, we once more apply the uninformative prior but this time on orientations: that the aerosolized particles do not have any orientation bias when injected into the path of the x-ray pulses (i.e. $p(Q)$ is a constant). 
Following this, we need to determine the most probable fluence rescaling factor for each pattern, $\phi_K$, as it has been demonstrated to be vital for accurate multiple model reconstruction \cite{daurerPtychographicWavefrontCharacterization2021}.
For this purpose, we conducted a single model EMC reconstruction \cite{lohReconstructionAlgorithmSingleparticle2009} on each dataset $K$ to ascertain the most probable rescaling factor $\widetilde{\phi}_K$ for each pattern \cite{ayyerDragonflyImplementationExpand2016}. Subsequently, we made the assumption that $p(\phi_K) = \delta(\phi_K - \widetilde{\phi}_K)$. The two assumptions outlined in this paragraph result in a streamlined version of the likelihood function in \cref{eq:pkrho}, which is employed to assign weight to model importance in our Monte Carlo scheme:
\begin{equation}
    p(K\given \rho) \propto \sum_{Q \in \mathbb{Q}} p(K\given Q,W, \widetilde{\phi}_K)\; .
    \label{eq:pkrho2}
\end{equation} 

\subsection{Four-ball artificial model}\label{subsec:abg}
As shown in \cref{fig:framework}(c), the artificial model consists of four identical balls with centers at $\mathbf{0}$, $\mathbf{a}_1$, $\mathbf{a}_1 + \mathbf{a}_2$, and
$\mathbf{a}_1 + \mathbf{a}_2 + \mathbf{a}_3$.
The diameters of these balls are of unit length. 
In other words, 
$\envert{\mathbf{a}_1} = \envert{\mathbf{a}_2} =\envert{\mathbf{a}_3}=1$.
We are only concerned with the model's structure, which is described by three degrees of freedom: $\alpha$, $\beta$ and $\gamma$. These are chosen as follows:
\begin{align*}
    \alpha &= \langle -\mathbf{a}_1, \mathbf{a}_2\rangle\text{,} \\
    \beta &= \langle -\mathbf{a}_2, \mathbf{a}_3\rangle \text{,} \\
\gamma &= \langle \mathbf{a}_1 \times \mathbf{a}_2, \mathbf{a}_2 \times \mathbf{a}_3\rangle
\text{,} \\
\end{align*}
where $\langle \mathbf{v}_1, \mathbf{v}_2\rangle$ represents the angle between vectors $\mathbf{v}_1$ and $\mathbf{v}_2$.
To generate new models, we perturb our canonical four-ball model, $(\alpha, \beta, \gamma) \to (\alpha +\delta_1, \beta+\delta_2, \gamma+\delta_3)$, where $\delta_{i=1,2,3} \in [ -0.02, 0.02]$ are three independent uniform random numbers.
We impose an extra constraint on the perturbed models that their individual $\alpha, \beta, \gamma$ cannot exceed the range $[1, 2.5] \times [1.2, 2.0] \times [0, 2.0]$. 
Perturbations that violate this constraint are discarded.

To generate the \num{100000} diffraction patterns (i.e., upper black line in \cref{fig:framework}(d), we first generated an ensemble of 100 perturbed models within the constrained angular ranges in the previous paragraph.
Then \num{1000} diffraction patterns were generated from each perturbed model at random 3D orientations.

In the Monte Carlo search, to average out the effect from the choice of first sampled model, we sampled the dataset with 12 different random initial models, resulting 12 trajectories. The each trajectory has a length of \num{5000}.

\subsection{Strategies to accelerate Markov Chain Monte Carlo}

The importance weight of each candidate model $\rho$ is linked to the model's data likelihood $p(K\given \rho)$ (\cref{eqn:likelihood}). 
We accelerated this calculation with the following four strategies.  

First, we partitioned the Monte Carlo model searches into smaller searches performed in parallel.
Each dataset of diffraction patterns, $\bbK_\text{oct30}$ or $\bbK_\text{oct40}$, was randomly split into five similarly-sized, non-overlapping partitions. 
We accumulated eight different pools of Monte Carlo models for each partition, each containing \num{5000} models.
Each pool was started from a randomly perturbed version of the same average model.  
Eventually, we accumulated \num{400000} models: \num{200000} for $\bbK_\text{oct30}$, and \num{200000} for $\bbK_\text{oct40}$.

Second, the determination of every single diffraction pattern's orientation with respect to each 3D model is performed only once -- when the 3D model is first added to the model pool. 
When additional models are added to this pool, we only need to rescale existing models' weights without comparing the latter against the diffraction data again.
Overall, the number of orientations to be determined scaled like the product of the number of models and number of diffraction patterns.

Third, we accelerated the calculation of the likelihood $p(K\given \rho)$, which as defined in \cref{eq:pkrho}, compares each pattern $K$ against tomograms of all possible orientations of each 3D FFTO model $\rho$ in the model pool. 
However, in practice, only the likelihoods of a few orientations within each model were significant \cite{shenEncryptionDecryptionFramework2021}. Put differently, $p(K\given\rho)$ is sparse. 
Hence, we used coarse orientation sampling to first identify rotational neighbourhoods near these significant orientations.
Then we increased the orientation sampling around these neighbourhoods for each data-model pair ($K$ and $\rho$).

Fourth, we employed a memory-efficient approach to compute the two-dimensional Ewald sphere intensity section of each model $\rho$. 
A direct way to perform this job is to voxelize the real-space electron density of $\rho$, and then apply a fast Fourier transform on this. 
Instead, since $\rho$ is a polyhedron with uniform density, we can compute its Fourier transform more accurately using a finite-element approach (see \nameref{sec:Method}).
Briefly, each polyhedron is partitioned into non-overlapping tetrahedra, whose separately complex-valued Fourier transforms can be analytically computed and then coherently added together to give the Ewald sphere section of the original polyhedron.

These four time- and memory-saving approaches were implemented across 20 parallel-running NVIDIA GTX 1080 Ti GPUs.
The \num{400000} models for the $\bbK_\text{oct30}$ and $\bbK_\text{oct40}$ datasets were accumulated in approximately 240 hours.

\subsection{{\it In silico} filtration with 2D EMC}

\begin{figure}
    \centering
    \includegraphics{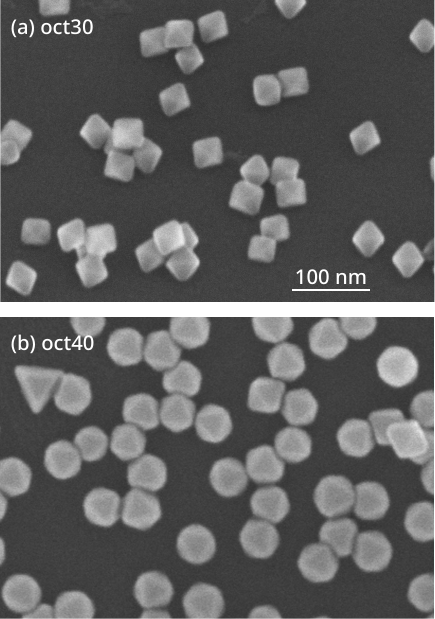}
    \caption{(a) and (b) represent SEM images of the oct30 and oct40 nanoparticle samples, respectively, each exhibiting nominal average widths of \SI{30}{nm} and \SI{40}{nm}. It is important to acknowledge that within the original sample, not all particles adopt an octahedral shape. Nevertheless, these non-octahedral variants can be effectively distinguished and filtered out through the application of the 2D EMC classification process.
}
    \label{fig:sem}
\end{figure}

\begin{figure}
    \centering
    \includegraphics{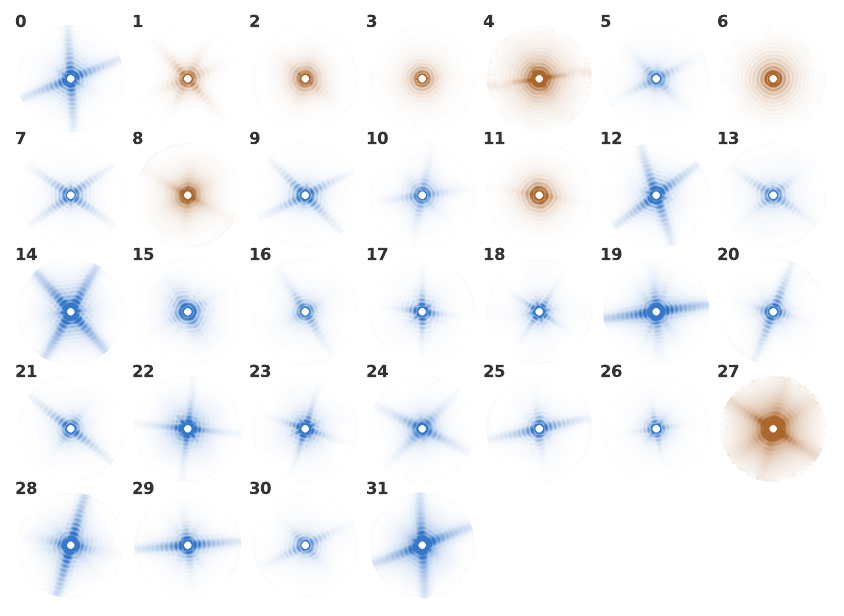}
    \caption{This is the 2D EMC classification of the raw oct40 dataset. Clusters colored with dark red are unlikely generated from an octahedral sample, hence filtered out for study in this paper.}
    \label{fig:2demc}
\end{figure}

As SEM images in \cref{fig:sem} show, our synthesized particles contained shapes that did not resemble truncated octahedra.
Similar to previous analyses of this EuXFEL dataset \cite{ayyer3DDiffractiveImaging2021} , we filtered out ({\it in silico}) some of the undesirable data heterogeneities using 2D classification via EMC method\cite{ayyer3DDiffractiveImaging2021}.
This method classifies diffraction patterns into multiple 2D models up to an overall in-plane rotation. 
This effectively helps us to identify significant patterns unlikely to arise from single truncated octahedra without having to reconstruct or compare them against 3D models.
In \cref{fig:2demc}, these non-conforming patterns (dark red) clusters include: multiple-particle shots or triangular particles (cluster 1, 2 and 8), spherical patterns (cluster 2, 3, 4, 6, 11 and 27; absence of prominent streaks), and patterns with feature-less stripe (cluster 4).
To increase the concentration of truncated octahedra, 2D EMC was applied in three rounds on the oct30 and oct40 datasets separately.
After each round, non-conforming clusters were manually identified (like these dark red clusters in \cref{fig:2demc}) and discarded before the next round.
Only patterns that survived all three times of filtration were used for this paper: \num{1282}k out of \num{1608}k for the oct30 dataset, and \num{823}k of \num{1032}k for the oct40 dataset.

For reference, we show typical 2D intensity slices of an ideal truncated octahedron from different orientations in the supplementary (\cref{fig:match}).

\subsection{Finite-element Fourier transform}\label{subsec:finite}
\begin{figure}[htpb]
    \centering
    \includegraphics{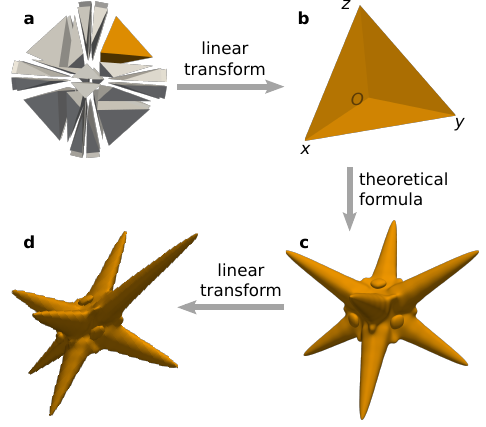}
    \caption{
        (a) A polyhedral volume is divided into several smaller non-overlapping tetrahedra. Each tetrahedron includes the coordinate origin $(0,0,0)$ and the three vertices of its triangulated meshes. (b) Each tetrahedra is linearly transformed to a standard trirectangular tetrahedron. (c) We can compute the complex-valued Fourier transform of each  suitably "rectangularized" tetrahedron in (b). Here we show contours of such Fourier intensities. (d) The linear transformation from (a) to (b) is be reversed to obtain the Fourier transform of the highlighted (in orange) tetrahedron in (a). 
    }
    \label{fig:finite}
\end{figure}

The finite-element Fourier transform method explicitly calculates the Fourier transform of a uniform density polyhedral volume parameterized by its surface vertices without initially 'voxelizing' the volume onto a grid.
A `voxelized' electron density is conventionally needed to compute its 3D discrete Fourier transform (DFT), which can be readily compared with its XFEL diffraction patterns. 
Here, the inter-particle variations of $L_\text{long}$ within the oct30 ensemble measures only 2-voxels in a 3D electron density array with size $251^3$ according to the Nyquist–Shannon sampling theorem
\cite{ayyerDragonflyImplementationExpand2016}. 
To minimize significant truncation errors when describing small but measurable size/shape variations among the nanoparticles using the voxelization approach, these volumes are typically padded with extra zeros (equivalent to oversampling their Fourier volumes by a multiplicative factor of $\alpha$).

As shown in \cref{fig:finite}, due to the linearity of the Fourier transform, the Fourier transform of a 3D volume is the sum of the Fourier transforms of its constituent non-overlapping tetrahedra.
We derive the analytical formula for computing the Fourier transform of arbitrary tetrahedra in the \nameref{sec:supplementry} material. 

The complexity of computing Fourier transforms using this tetrahedralization method is $O(NM)$, where $N$ is the number of point samples in the 3D Fourier volume and $M$ is the number of tetrahedra. 
For the truncated octahedra in this work, $M=48$.
Comparatively, the computational complexity of `voxelizing' then performing Fast Fourier Transform (FFT) on each unique polyhedra scales like $O(N \alpha \log_2 (\alpha N) )$, where $\alpha$ is the typical oversampling parameter needed to overcome the truncation issue with voxelizing polyhedra (discussed above).

Hence our tetrahedralization method will generally perform faster when $M < \alpha \log_2 (\alpha N)$. 
This is true when we coarse-grain our candidate nanoparticles to polyhedra with relatively few faces (small $M$) while our diffraction patterns are highly oversampled (large $\alpha N$). 
There are further time savings for the tetrahedralization method when we only need to compute a fraction of the full Fourier intensities (e.g., only along a handful of Ewald sphere slices).

\subsection{Free Facet Truncated Octahedron (FFTO)} \label{subsec:FFTO}
The FFTO model consists of 14 facets (\cref{fig:pca}(b)): six for the $(100)$ directions, and eight for the $(111)$ directions. 
Each facet is described by a 3D vector $\bvec h = (A, B, C)$, where  $(A, B, C)$ is a point on the facet where the vector $(A, B, C)$ is also normal to the facet. 
The plane equation for such a facet is $\bvec h \cdot (x, y, z) = |\bvec h|^2$ or
\begin{equation}
    Ax+By+Cz = A^2+B^2+C^2\text{.}
\end{equation}

In total, $42=14\times 3$ parameters are needed for each FFTO model.
For the special case of an ideal FFTO model with perfect octahedral symmetry, their facets are described by the six cyclic permutations of $(\pm a, 0, 0)$ that describe the (100) facets, plus the eight combinations of $(\pm b, \pm b, \pm b)$ for the (111) facets.

To perturb an FFTO model, each facet, $(A, B, C)$ is mapped to a new facet $(A, B, C)+\mathbf{v}$, where $\mathbf{v}$ is 3D uniform random vector within a \SI{0.84}{nm}-radius ball.
Each perturbation also needs to satisfy two constraints. 
The first constraint is that a model has to be a convex volume.
The second constraint ensures each FFTO model stays reasonably close to the ideal truncated octahedron.
To enforce these two constraints, the closest ideal truncated octahedron model (described in the previous paragraph) is found first for a given candidate FFTO model. 
This closeness is defined as the Euclidean distance between the 24 corresponding pairs of vertices between the two models.
Hence, the closest ideal truncated octahedron to a perturbed FFTO model minimizes this total distance between the two models. 
If the distance between any two paired vertices between these two models is larger than \SI{1.68}{nm}, then the perturbed FFTO candidate is rejected.
For the Monte Carlo importance sampling, we will continue to perturb each FFTO model until these two constraints are satisfied.

\subsection{Eliminate symmetry redundancy}\label{subsec:eli}
Here we explain how we checked if two FFTO models are similar up to a particular permutation of their facet indices.
This check is used to re-order the facet indices of our pool of models in \cref{fig:pca} to then distill model-model differences that are not due to trivial permutations of each model's facet indices.
Each FFTO model is uniquely represented by an 14-element vector that detailed the areas of each FFTO model's 14 facets.
Before any new facet-index permutation is attempted on model $\rho$, its 14-element area vector, $\mathbf{A}_\rho$, is normalized to $\mathbf{A}_\rho/V^{2/3}$ where $V$ is the model's volume.

Rather than checking and permuting all possible pairs of models in our pool $\mathbb{M}$, we aim to to re-order each model's facet index to have the smallest distortion from the pool's average area vector $\overline{\mathbf{A}} = \frac{1}{|\mathbb{M}|}\sum_{\rho \in \mathbb{M}} \mathbf{A}_\rho$.
This re-ordering is performed iteratively with two alternating steps: in the first step we compute $\overline{\mathbf{A}}$ given each model's current index order; then in the second step we re-order each model's indices to minimize the model's area vector from the mean vector using $\argmin_{\hat{P}} |\hat{P}(\mathbf{A}_\rho) - \overline{\mathbf{A}}|^2$ where $\hat{P}$ refers to the facet-index permutation over the symmetry orbit of the ideal truncated octahedra.
This iterative procedure is repeated until all re-ordering ceases and the mean vector stops changing. 

\subsection{Convergence of posterior estimation}\label{subsec:validation}
We need to determine if our Markov Chain Monte Carlo (MCMC) scheme has accumulated a sufficiently large pool of models $\bbM$ that adequately samples the posterior distribution over all possible models $\{ \rho\}$. 
Our demonstration of convergence comprises two steps. 
First, for our posterior estimation to have converged, it is {\it necessary} that the distribution differences between two iterations should be sufficiently small after convergence, or 
\begin{equation}
    \sum_\nu \bigl\Vert p(\nu\given\bbK; \bbM^{(n)})\cdot\nu - p(\nu\given\bbK;\bbM^{(m)})\cdot\nu\bigr\Vert\;,
    \label{eq:convergence}
\end{equation}
is a sufficiently small value when $m$, $n$ are sufficiently large, where $m$ and $n$ are iteration numbers, $\bbM^{(n)}$ is the sampled model pool at $n^\text{th}$ iteration, and some normalization, $\Vert\cdot\Vert$, is used here for multiple-dimensional $\nu$.  
Second, we further corroborate this convergence if $\bbM$ is a {\it self-consistent generative model}. 
We demonstrate this self-consistency by assuming a subset model pool $\bbM^\prime \in \bbM$ as the synthetic ground truth from which a number of diffraction data are generated $\bbK^\prime$; we then repeated our MCMC posterior estimation on $\bbK^\prime$ to obtain a third model pool, $\bbM^{\prime\prime}$. 
For our posterior estimation $\bbM$ to have converged, it is necessary that the posterior predictive $p_T(\nu|\mathbb{K})$ marginalized over $\bbM$, $\bbM^\prime$, and $\bbM^{\prime \prime}$ are sufficiently similar.

\begin{figure*}[htpb]
    \centering
    \includegraphics{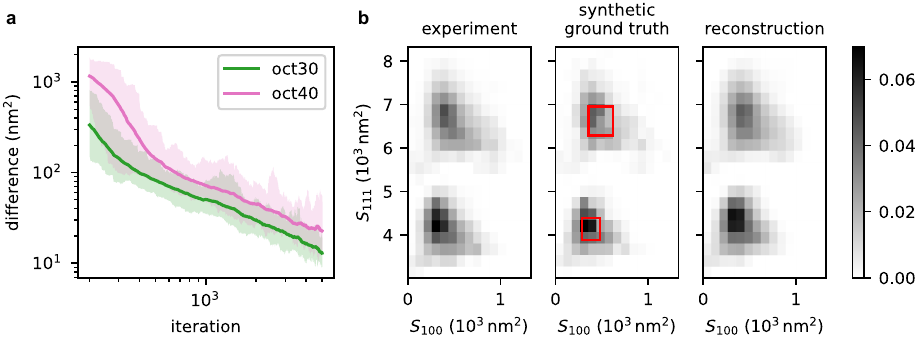}
    \caption{Validating the convergence and self-consistency of our reconstructed posterior predictive distribution $p(S_{111}, S_{100} \given \bbK)$. 
    (a) {\it Convergence} in the difference areas between the most likely models in our reconstructed model pools from the $n$-th and $(n+200)$-th MCMC iteration. 
    The maximum and minimum changes in areal differences averaged over 40 trajectories shown as faint color fills around the mean change (darker line). 
    (b) {\it Self-consistency} in our reconstructed $p(S_{111}, S_{100} \given \bbK)$. Left panel (experiment): computed from Bayes model averaging over the pool of FFTO models $\bbM$ given the diffraction data $\bbK$. 
    Middle panel (synthetic ground truth): $p(S_{111}, S_{100} \given \bbK)$ of a synthetic ground truth model ensemble generated from a random subset of $\bbM$. 
    Right panel (reconstructed): $p(S_{111}, S_{100} \given \bbK)$ of a new pool of models $\bbM^\prime$ reconstructed from random patterns generated by the synthetic ground truth.}
    \label{fig:simulation_dist}
\end{figure*}

\Cref{fig:simulation_dist}(a) shows the convergence of our posterior predictive distribution $p(S_{111}, S_{100} \given \bbK)$, where the feature pair $\nu = \{S_{100}, S_{111}\}$ are the total areas of each FFTO model's $(100)$ and $(111)$ facets respectively.
In practice, as most patterns are only in favor of one model, to speed up the calculation, we count only the best matched model for each pattern instead of strictly following the definitions in \cref{eq:pnubbKbbM,eq:convergence}.
We denote the area difference between two models, $\rho_a$ and $\rho_b$, as $d(\rho_a, \rho_b)=d_{111}(\rho_a, \rho_b) + d_{100}(\rho_a, \rho_b)=\envert[1]{S_{111}(\rho_a) - S_{111}(\rho_b)} + \envert[1]{S_{100}(\rho_a) - S_{100}(\rho_b)}$. Then \cref{fig:simulation_dist}(a) summarizes the change in area between the $n^{\text{th}}$ and $(n+200)^{\text{th}}$ MCMC iteration as $\bigl\langle d\bigl(\rho^{(K)}_n,\rho^{(K)}_{n+200}\bigr)\bigr\rangle_{K\in\bbK}$, where the $\rho_n^{(K)}$ stands for the best matched model for a pattern $K$ in the model pool $\bigl\{\rho_1, \rho_2, \dots, \rho_n\bigr\}$.
The colored fills in \cref{fig:simulation_dist}(a) span the largest and smallest values among all 40 trajectories (eight trajectories for all
five non-overlapping partitions of the full dataset) at each iteration. 
By iteration $n=5000$, the magnitude of this areal differences is about \SI{10}{nm^2}, which is less than $1\%$ compared to the total area of a
particle.

In the second step of our validation, we tested for self-consistency of the MCMC model pool $\bbM$ that was reconstructed from diffraction data $\bbK$.
From $\bbM$ we picked the \num{2000} best-matched models of \num{2000} randomly selected patterns in $\bbK$. 
These \num{2000} models forms a {\it synthetic ground truth} pool of models $\bbM^\prime$.
We then generated \num{1000} diffraction patterns from each model in $\bbM^\prime$, denoting these patterns as $\bbK^\prime$.
Each of these patterns are randomly oriented, and rescaled from the distribution of factors recovered in the earlier single-model EMC reconstruction of $\bbK$ that initialized the reconstruction of $\bbM$ \cite{ayyerDragonflyImplementationExpand2016}.
Thereafter, we used the same MCMC procedure used to recover a third model pool $\bbM^{\prime \prime}$ from $\bbK^\prime$.
\Cref{fig:simulation_dist}(b) shows three posterior predictive distributions marginalized over the model pool $\bbM$ reconstructed from $\bbK$, the synthetic ground truth $\bbM^\prime$, and $\bbM^{\prime\prime}$ reconstructed from $\bbK^\prime$. 
Since we know the ground truth models $\bbM^\prime$ for every pattern in the synthetic dataset $\bbK^\prime$, we can evaluate the area differences, $d_{111}$ and $d_{100}$, between the ground truth models and reconstructed best-matched models in $\bbM^{\prime \prime}$.
The two red rectangles in \cref{fig:simulation_dist}(b) mark the average difference in $d_{111}$ and $d_{100}$ for the oct30 and oct40 datasets.

\subsection{PDDA coverage} \label{subsec:coverage}
\begin{figure}[htpb]
    \centering
    \includegraphics{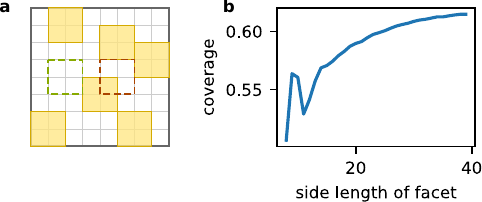}
    \caption{Fractional coverage of non-overlapping PDDA molecules on a crystal facet depends on the facet's area. 
    (a) The grey lattice represents a crystal facet covered by PDDA molecules (yellow squares). The PDDA molecules are not allowed to overlap (i.e., a next PDDA can occupy the green dashed square but not the red one).
    (b) The PDDA coverage vs.~the side length of a facet.}%
    \label{fig:coverage}
\end{figure}

A simple model is proposed to show the finite size effect on PDDA coverage on crystal facet growth at elevated temperatures
(\cref{fig:coverage}(b)). 
The synthesis protocol creates different shaped Au nanoparticles by adding PDDA polymer chains to the growth solution \cite{liFacilePolyolRoute2008,luSizetunableUniformGold2017}.  
Each PDDA polymeric molecule has probability of attaching to the crystal facets only if there is sufficient areal contact between them. 
In this model, we use an $N\times N$
square lattice (the gray lattice in \cref{fig:coverage}(a)) to simulate a crystal facet. 
Thus $N$ could be regarded as the side length of a crystal facet whose typical size is few tens nanometers.
The PDDA molecules that attach to the facet are abstracted as an $L\times L$ square (yellow square in \cref{fig:coverage}(a)). 
We attempt to place PDDA molecules randomly over this facet such that no two PDDA molecules overlap. 
This mutual exclusion requirement expresses 

Then the coverage is $n L^2 / N^2$, where $n$ is the number of PDDAs placed. Since the size of a PDDA is few nanometers, we choose $L=4$ in the simulation. For each $N$, simulations were run \num{20000} times. 
As shown in \cref{fig:coverage}(b), the average coverage is increasing with the side length of a
facet, which causes effectively smaller surface tension.

\section{Supplementary}\label{sec:supplementry}
\subsection{Match 2D EMC clusters with ideal 2D intensity slices}
\begin{figure}
    \centering
    \includegraphics{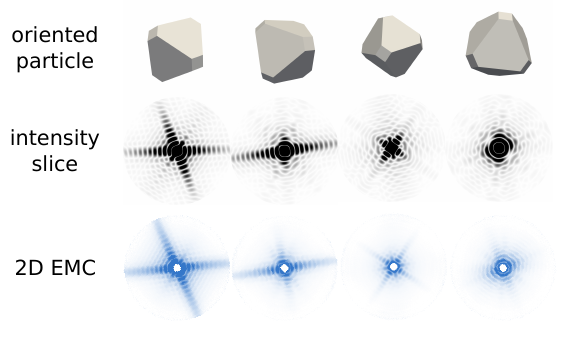}
    \caption{Here, we select four typical 2D EMC clusters (blue). For each one, a 2D intensity (black) slice is manually matched, and the corresponding oriented particle is showed in grey.}
    \label{fig:match}
\end{figure}

In the \cref{fig:match}, we manually match several typical 2D EMC clusters with 2D intensity slices of an ideal octahedra from different orientations. The shape of this octahedra is given by the average model we reconstructed.
\subsection{Finite volume Fourier Transformation}
\label{subsec:FiniteFT}
Comparing to biomolecules, nanoparticles are also a common kind of sample in XFEL SPI experiment but has a much simpler structure.
Usually, we could assume its density is uniform and has a polyhedron
shape. In this subsection, we propose a numerical Fourier transformation scheme for any uniform
polyhedron $S$. This method could avoid the voxelization of $S$. As the difficulty of the Fourier transformation of $S$ , $\iiint_S
\exp(\img \bvec k \cdot \bvec x)\,\dif \bvec x$
comes from the complexity of the shape of $S$,
we convert this one big integral into several smaller integrals over tetrahedrons.
The surface (boundary) of $S$, $\partial S$, can be triangulated into a list of $n$
triangular faces
\[
\bigl\{ [v_1^{(i)}, v_2^{(i)}, v_3^{(i)}] \given[\big] i=1, 2, \dots, n \bigr\}\text{,}
\] 
where $v^{(i)}$s are the vertices of triangular $i$, $[v_1^{(i)}, v_2^{(i)}, v_3^{(i)}]$.
Suppose $S$ is convex and the origin $v_O$ is inside $S$. Then $S= \sum_i [v_1^{(i)}, v_2^{(i)}, v_3^{(i)}, v_O]$
where $[v_1^{(i)}, v_2^{(i)}, v_3^{(i)}, v_O]$ is the tetrahedron with base
$[v_1^{(i)}, v_2^{(i)}, v_3^{(i)}]$ and apex $v_O$. Immediately, we get
\begin{equation}
    \iiint_S \exp(\img \bvec k \cdot \bvec x)\,\dif \bvec x
    = \iiint\limits_{\sum_i [v_1^{(i)}, v_2^{(i)}, v_3^{(i)}, v_O]} \exp(\img \bvec k \cdot \bvec x)\,\dif \bvec x\text{.}
    \label{eq:splitS}
\end{equation}

\Cref{eq:splitS} retains its validity even when these two assumptions are relaxed. A straightforward argument in support of this is that both sides of \cref{eq:splitS} exhibit continuity across all vertices, and the integral remains independent of the choice of origin. The critical factor here is ensuring the correct orientation of a surface, which is determined by the sequential selection of three vertices, $\{v_1^{(i)}, v_2^{(i)}, v_3^{(i)}\}$, on the surface following the right-hand rule. It is imperative that the orientation, represented by the ``thumb'' direction, points outward from the $S$. A more rigorous mathematical statement pertaining to this concept, ``orientability'', can be found in most algebraic topology textbooks.

The region covered by $[v_1^{(i)}, v_2^{(i)}, v_3^{(i)}, v_O]$ and 
$[v_x, v_y, v_z, v_O]$ can be converted into each other by a linear transform, $A$,
\begin{equation}
    A^{-\text{T}}[\bvec v_1, \bvec v_2, \bvec v_3, \bvec v_O] = 
    \begin{bmatrix}
        1 & &  & 0\\
          & 1 & & 0 \\
          & & 1 & 0
    \end{bmatrix}
    = [\bvec v_x, \bvec v_y, \bvec v_z, \bvec v_O]
\end{equation}
where 
\begin{equation}
    A = 
    \begin{bmatrix}
       \bvec v_1^\text{T} \\ 
       \bvec v_2^\text{T} \\ 
       \bvec v_3^\text{T}
   \end{bmatrix}\text{,}
\end{equation}
$v_x$, $v_y$, and $v_z$ are three unit points of $x$, $y$, and $z$-axis, and
the $\bvec v$ emphasizes that it is a column vector of vertex $v$. 
The integrals over $[v_1^{(i)}, v_2^{(i)}, v_3^{(i)}, v_O]$ in \cref{eq:splitS}
can be converted to integrals over the same trirectangular triangular pyramid
$[v_x, v_y, v_z, v_O]$.
\begin{align*}
    \iiint_S \exp(\img \bvec k \cdot \bvec x)\,\dif \bvec x 
    &= \sum_i \iiint\limits_{[v_1^{(i)}, v_2^{(i)}, v_3^{(i)}, v_O]} \exp(\img \bvec k \cdot \bvec x)\,\dif \bvec x\\
    &= \sum_i \det A^{(i)}\iiint\limits_{[v_x, v_y, v_z, v_O]} \exp(\img A^{(i)}\bvec k \cdot\bvec x)\,\dif \bvec x\\
    &= \sum_i \det A^{(i)} \cdot F^\star(A^{(i)}\bvec k)
\end{align*}
where $F^\star$ is the Fourier transformation of $[v_x, v_y, v_z, v_O]$. 
The closed-form expression for $F^\star$ can be found by
\begin{align*}
    F^\star(\bvec k)
    &= \int_0^1\int_0^{1-x}\int_0^{1-x-y}\exp\big[\img(k_x x + k_y y + k_z z)\big]\,\dif z\dif y \dif x \\
    &= \dfrac{\img \sum_{xyz}\big[\exp(\img k_x)-1\big]k_y k_z (k_z - k_y)}{\prod_{xyz}k_x\prod_{xyz}(k_x-k_y)}\text{,}
    \numberthis\label{eq:Fstar}
\end{align*}
where $\sum_{xyz}$ and $\prod_{xyz}$ are the index-rolling summation and production,
for instance, $\sum_{xyz}k_x = k_x + k_y + k_z$, $\prod_{xyz}(k_y - k_z) = (k_y - k_z)(k_z - k_x)(k_x - k_y)$.

In the numerical calculation of \cref{eq:Fstar}, zero denominator in \cref{eq:Fstar}
causes zero division error. Those cases are supposed to be handled separately
by calculating the limits of \cref{eq:Fstar}.
It should be pointed out that as the right-triangular pyramid has $\mathrm{C}_{3\mathrm{v}}$ symmetry,
all permutations $(\sigma(x)\ \sigma(y)\ \sigma(z))$, like $(y\ x\ z)$, to the index list $(x\ y\ z)$ in $F^\star(k_x, k_y, k_z)$
give the same values,
which also could be verified directly from \cref{eq:Fstar}.
Therefore, all zero-division cases are divided into six classes, and the condition label for
each class is a representative of its index-permutated class.
\begin{enumerate}
    \item $k_x=k_y=k_z=0$
        \[F^\star(0, 0, 0)=\frac{1}{6}\text{;}\]
    \item $k_x=k_y=k_z\neq 0$
        \[F^\star(k_x, k_x, k_x)=\dfrac{-2\img+\exp(\img k_x)(2\img + 2k_x -\img k_x^2)}{2k_x^3}\text{;}\]
    \item $k_x \neq 0$, $k_y = k_z = 0$
        \[F^\star(k_x, 0, 0)=\frac{1}{k_x^2}+\frac{\img\left[ -2 + 2\exp(\img k_x)+k_x^2 \right]}{2k_x^3}\text{;}\]
    \item $k_z=k_x\neq 0$, $k_y=0$
        \[
            F^\star(k_x, 0, k_x) = -\dfrac{-2\img+k_x +\exp(\img k_x)(2\img +k_x)}{k_x^3}\text{;}
        \]
    \item $k_x=0$, $k_y\neq 0$, $k_z\neq 0$, $k_y\neq k_z$
        \[
            F^\star(0, k_y, k_z) = \dfrac{\img\big[\exp(\img k_y)-1 \big]}{k_y^2(k_y-k_z)}+
            \dfrac{\img\big[\exp(\img k_z)-1 \big]}{k_z^2(k_z-k_y)} -
            \frac{1}{k_y k_z}\text{;}
        \]
    \item $k_x = k_y \neq 0$, $k_z \neq 0$
        \begin{align*}
            F^\star(k_x, k_x, k_z) = \dfrac{1}{k_y^2(k_y-k_z)^2k_z}
            \Big\{&\img \big[\exp(\img k_y)-1\big]k_z^2+\\
                  & \img k_y^2\big[\exp(\img k_z)-1+\img\exp(\img k_y)k_z\big] + \\
                  & k_y k_z\big[2\img +\exp(\img k_y)(-2\img + k_z)\big]\Big\}\text{.}
        \end{align*}
\end{enumerate}

\bibliography{ref}

\end{document}